\documentclass[fleqn,usenatbib]{mnras}

\usepackage{newtxtext,newtxmath}

\usepackage[T1]{fontenc}

\DeclareRobustCommand{\VAN}[3]{#2}
\let\VANthebibliography\thebibliography
\def\thebibliography{\DeclareRobustCommand{\VAN}[3]{##3}\VANthebibliography}

\usepackage{graphicx}	
\usepackage{amsmath}	
\usepackage{float}      

\title[Pisces VII]{Pisces VII: Discovery of a possible satellite of Messier 33 in the DESI Legacy Imaging Surveys}

\author[Martinez Delgado et al.]{David Mart{\'\i}nez-Delgado$^{1}$\thanks{Talentia Senior Fellow},
Noushin Karim$^{2}$, Emily J. E. Charles$^{2}$,
Walter Boschin$^{3}$ $^{4}$ $^{5}$,
Matteo Monelli $^{4}$ $^{5}$,
\newauthor Michelle L. M. Collins $^{2}$,
Giuseppe Donatiello $^{6}$,
Emilio J. Alfaro $^{1}$\\
$^{1}$Instituto de Astrof\'isica de Andaluc\'ia, CSIC, Glorieta de la Astronom\'\i a, E-18080, Granada, Spain \\
$^{2}$Physics Department, University of Surrey, Guildford, GU2 7XH, UK\\
$^{3}$Fundaci\'on G. Galilei - INAF (Telescopio Nazionale Galileo), Rambla J. A. Fern\'andez P\'erez 7, E-38712 Bre\~na Baja (La Palma), Spain\\
$^{4}$Instituto de Astrof\'isica de Canarias (IAC), Calle V\'ia L\'actea s/n, E-38205 La Laguna, Tenerife; Spain \\
$^{5}$Facultad de F\'isica, Universidad de La Laguna, Avda. Astrof\'isico Fco. S\'anchez s/n, 38200La Laguna, Tenerife, Spain. \\
$^{6}$UAI -- Unione Astrofili Italiani /P.I. Sezione Nazionale di Ricerca Profondo Cielo, 72024 Oria, Italy \\
}

\pubyear{2021}

\begin{document}
\label{firstpage}
\pagerange{\pageref{firstpage}--\pageref{lastpage}}
\maketitle
\begin{abstract}
We report deep imaging observations with DOLoRes@TNG of an ultra-faint dwarf satellite candidate of the Triangulum galaxy (M33) found by visual inspection of the public imaging data release of the DESI Legacy Imaging Surveys. Pisces VII/Triangulum (Tri) III is found at a projected distance of $72\,{\rm kpc}$ from M33, and using the tip of the red giant branch method we estimate a distance of $D=1.0^{+0.3}_{-0.2}\,{\rm Mpc}$, meaning the galaxy could either be an isolated ultra-faint or the second known satellite of M33. We estimate an absolute magnitude of $M_V=-6.1\pm0.2$ if Pisces VII/Tri II is at the distance of M33, or as bright as $M_V=-6.8\pm0.2$ if the galaxy is isolated. At the isolated distance, it has a physical half-light radius of $r_{\rm h}=131\pm61\,{\rm pc}$ consistent with similarly faint galaxies around the Milky Way. As the tip of the red giant branch is sparsely populated, constraining a precision distance is not possible, but if Pisces VII/Tri III can be confirmed as a true satellite of M33 it is a significant finding. With only one potential satellite detected around M33 previously (Andromeda XXII/Tri I), it lacks a significant satellite population in stark contrast to the similarly massive Large Magellanic Cloud. The detection of more satellites in the outskirts of M33 could help to better
illuminate if this discrepancy between expectation and observations
is due to a poor understanding of the galaxy formation process, or if it is due to the low luminosity and surface brightness of the M33 satellite population which has thus far fallen below the detection limits of previous surveys. If it is truly isolated, it would be the faintest known field dwarf detected to date.
\end{abstract}


\section{Introduction}

With a stellar mass of $M_{*}=3\times 10^9 M_{\odot}$ \citep{mcconnachie12}, and a halo mass
of $\sim 10^{11} M_{\odot}$ \citep{corbelli14}, the Triangulum spiral galaxy (M33) is the
most massive satellite of the Andromeda galaxy (M31) and one of the
most massive galaxies of the Local Group. With this mass, the
$\Lambda$CDM cosmological paradigm predicts that M33 should host a
number of its own satellites. Cosmological simulations find it should
have between 9-25 companions with stellar mass $>10^3 M_{\odot}$
(e.g. \citealt{dooley17,patel18,bose18}), and at least 1 with $M_{*}>10^6 M_{\odot}$ \citep{deason13}. However, to date, only one potential satellite has been
uncovered: Andromeda (And) XXII/Triangulum (Tri) I, which has a stellar mass of 2$\times 10^4
M_{\odot}$ \citep{martin09,martin16}. This was found in the framework of
the Pan-Andromeda Archaeological Survey (PAndAS; \citealt{mcconnachie09}), an
observing program which has conducted a deep survey of the M33 halo
out to $\sim$1/3 of its virial radius.\\

To date, no brighter satellites have been found outside this radius in shallower surveys such as the SDSS. There is also a notable dearth of globular clusters around M33 compared to other spiral galaxies \citep{cockcroft11}. This negligible satellite
population is in stark contrast to that of the similarly massive
Large Magellanic Cloud (LMC), which has upwards of 7 known satellite
galaxies (e.g. \citealt{jethwa16,fritz19,erkal20a}). Part of this is due to the difference in
limiting magnitude that can be probed in M33 versus the far nearer
LMC. But even so, the lack of bright companions ($M_{*}>10^4
M_{\odot}$) is surprising.\\

Previously, this lack of satellite galaxies and far flung globular
clusters of M33 was attributed to its dynamical evolution. Warps to
its outer stellar and HI disks were thought to originate from a prior interaction with M31 occurring $\sim$2 Gyr ago, which would have stripped much of M33's stellar halo and satellites \citep{mcconnachie09,cockcroft11}. Nevertheless, newer studies which include up-to-date proper motions for M31 and M33 suggest that the latter is more likely on its first infall to the M31 system (e.g. \citealt{patel17,vandermarel19}). If this scenario is correct, M33's satellite system should extend beyond a single, low mass satellite.\\

This dramatic gap between theoretical expectations and observations
could imply our understanding of the formation of low mass galaxies is flawed. Perhaps the feedback recipes used in hydro-dynamical surveys are incorrect, or we are wrong about the nature of dark matter itself. Or, it could be that the majority of M33 satellites have luminosities and surface brightnesses that lie just below the
detection limits of previous surveys. In any event, given the paucity of known M33 satellites, even a single new detection or exclusion of a companion has the potential to completely change our understanding of the M33 system, and galaxy formation more widely.

\begin{figure*}
	\includegraphics[width=\textwidth]{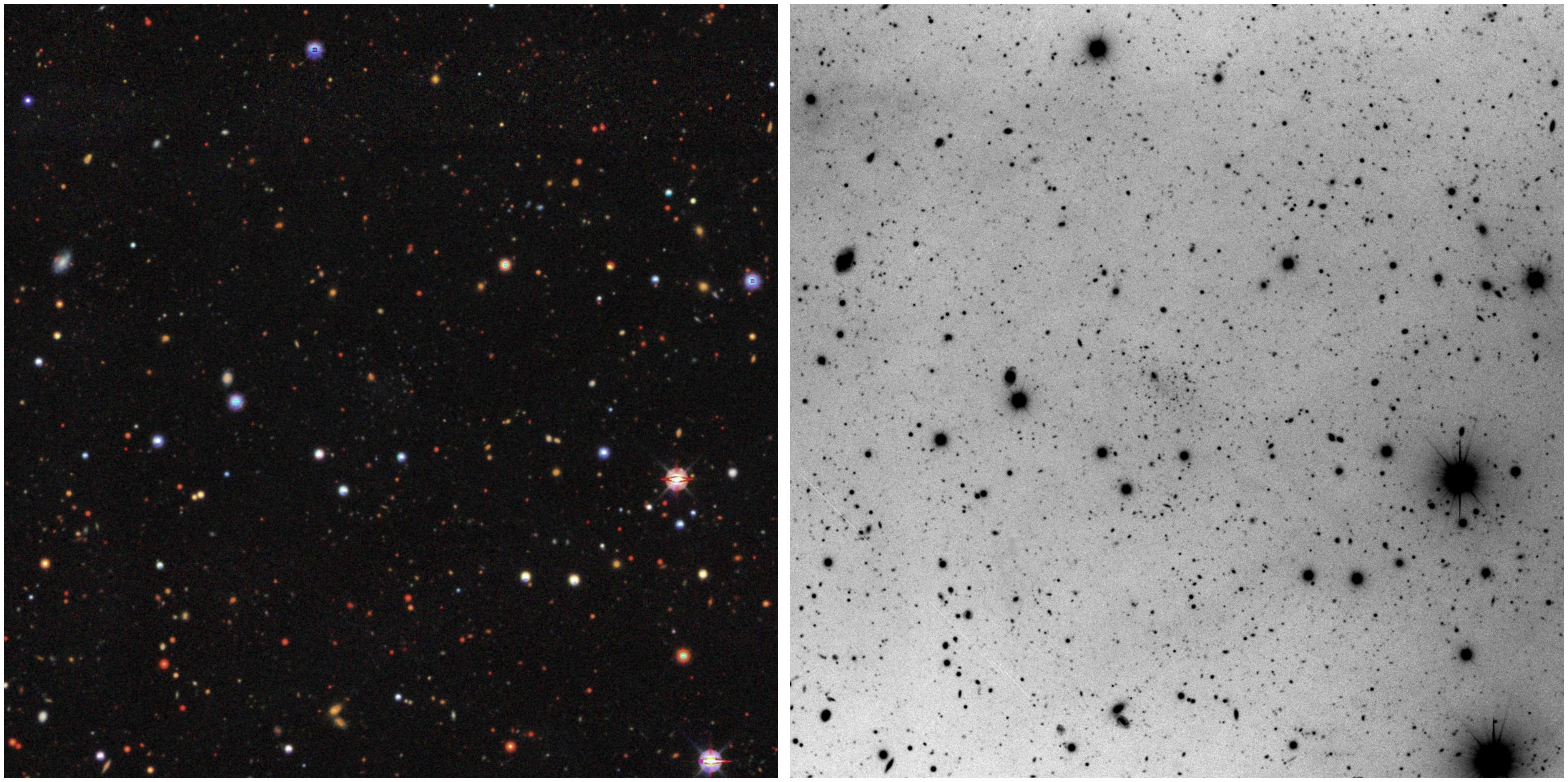}
    \caption{{\it Left panel}: Image of the dwarf galaxy Pisces VI/Tri III from the DESI LIS. {\it Right panel}: TNG $r$-band image of the galaxy obtained from TNG follow-up observations (see Sec. 2). The total field of view of both images is 8.6$\arcmin$ $\times$ 8.6$\arcmin$. North is up, East is left.}
    \label{fig:TNG}
\end{figure*}

In the last decades, the discoveries of Andromeda satellites have been made by means of visual inspection or 
automatic algorithm searches in stellar density maps of resolved red giant branch (RGB) stars, counted in selected 
areas of the color-magnitude  diagrams (CMDs) from large scale photometric survey data, such as the Sloan Digital Sky Survey (SDSS, \citealt{abazajian09}) and the Panoramic Survey Telescope and Rapid Response System (Pan-STARRs; \citealt{chambers16,martin13a,martin13b}). However, the main contribution to the dwarf census of M31 satellites came from 
PAndAS with the wide-field 
imager on the Canada French Hawaii Telescope (CFHT) \citep{mcconnachie09,martin09,richardson11}.

 Although the M31 stellar halo photometry from the PAndAS is significantly deeper than those from the SDSS or Pan-STARRs, this ground-based data can only reach the red clump locus in the CMDs at the distance of Andromeda. Thus, low-mass systems with absolute magnitude fainter than $M_{V}\sim -6$ are very hard to detect because of the lack of enough RGB star tracers in their CMDs, yielding a cut-off in the luminosity  function of satellites of M31 (see \citealt{brasseur11}; their Fig. 1). An alternative approach is to search for partially resolved stellar over-densities in the public deep images recently available from the  DESI Legacy Imaging Surveys  (DESI LIS; \citealt{Dey2019}). These data were obtained with the {\it Dark Energy Camera} (DECam) mounted on the Blanco 4-m telescope, located at the Cerro Torrolo Inter-American Observatory \citep{flaugher15}, which reach surface brightness as faint as $\sim$ 29 mag arcsec$^{\mathrm{-2}}$. This low surface regime allows us to detect the underlying, unresolved population of these systems as a diffuse light round over-density  overlapping a small ($\sim$ 1-2 arcmin) clump of faint stars.

Using this  approach, we have very recently identified a partial resolved dwarf galaxy candidate outside the PAndAS survey
footprint. Placed at the distance of M33, it would be located at a projected distance of $\gtrsim$70 kpc from the Triangulum galaxy. Unfortunately, the DESI LIS photometry is too shallow to allow the measurement of an accurate distance, making it impossible to reject it as a background isolated dwarf situated a few Mpcs behind Andromeda (as e.g. the Do~I dwarf; \citealt{martinezdelgado18}). In this paper we present follow-up photometric observations, a distance estimate and structural analysis of this stellar system that suggests its likely association to M33.

\section{Observations and data reduction}

Pisces ~VII/Tri III\footnote{Following the naming convention suggested in
the Appendix of Martin et al. (2009), we suggest for this galaxy the dual naming introduced with the discovery of Pegasus/And VI, Cassiopeia/And VII or Pisces VI/And XXII.} was discovered in the Pisces constellation by the amateur astronomer Giuseppe Donatiello as a partially resolved over-density in the Pisces constellation.  It was found by visual inspection of the available DESI LIS images of an extensive area of $20^{\circ}\times 30^{\circ}$ in the surroundings of the Triangulum galaxy (M33), outside the PAndAs footprint (see Fig. 1 left panel)\footnote{ Our search for dwarf satellite candidates around M33 was performed using the {\it Legacy Survey Sky Viewer} web site available in legacysurvey.org/viewer.}. The detection was subsequently confirmed by a visual inspection of the SDSS DR9 images and follow-up observations using the Italian Telescopio Nazionale Galileo (TNG) described in Sec. 2.1 . The position of the center of this new dwarf galaxy is given in Table~1.

\subsection{TNG imaging observations}
We used deep images of a 8.6'x8.6' field around the candidate galaxy
obtained with the focal reducer instrument DOLoRes (see
http://www.tng.iac.es/instruments/lrs/) of the 3.58-m TNG taken on November 17 2020 (program A42DDT2; P.I.: W. Boschin).

These observations include 41x180 sec unbinned (scale $0.252^{\prime\prime}$/pixel)
exposures in the g'-band and 20x180 sec unbinned exposures in the
r'-band, with a median seeing of ~1.15" and ~0.85", respectively.

The raw data were preprocessed in a routine way using standard IRAF
tasks, i.e. dividing the trimmed and bias-subtracted images by a
master flat field produced from multiple twilight sky-flat
exposures. 

Images were reduced using the DAOPHOT/ALLFRAME suite of programmes \citep{stetson87,
stetson94}, largely following the method outlined by \citet{monelli10b}. For each 
individual image, we performed the initial steps: i) search for stellar sources,
ii) aperture photometry, iii) PSF derivation, iv) PSF photometry with ALLSTAR.
Then images were registered with DAOMASTER and stacked on a median image. This
was used to extract a deeper list of stars, which was fed to ALLFRAME.
This provides individual catalogues with better determined position and instrumental
magnitude of the input sources. The updated photometry allowed us to refine the PSF
(using improved list of stars) and the geometric transformations (leading to a better coadded image
and cleaner input list). A final run of ALLFRAME provided the final photometry.
The list of sources was cleaned using the \textit{sharpness} parameter, and further polished by removing object after a visual inspection on the stacked image.

The photometric calibration was performed using local standard from the PanSTARRs 1 survey
\citep{chambers16}, using 121 stars in common. The mean magnitudes were
calibrated with a linear relation for the $g$ band, and a zero point for the $r$ band.
The calibrated local standard agree well below 0.01 mag with the tabulated ones,
with standard deviation of the order of 0.03 mag. All data are then extinction corrected using the reddening maps from \citet{schlafly11}.

\section{Methods}

\subsection{Identifying Probable Members}
\label{sec:Identifying Probable Members} 
\begin{figure*}
	\includegraphics[width=\columnwidth]{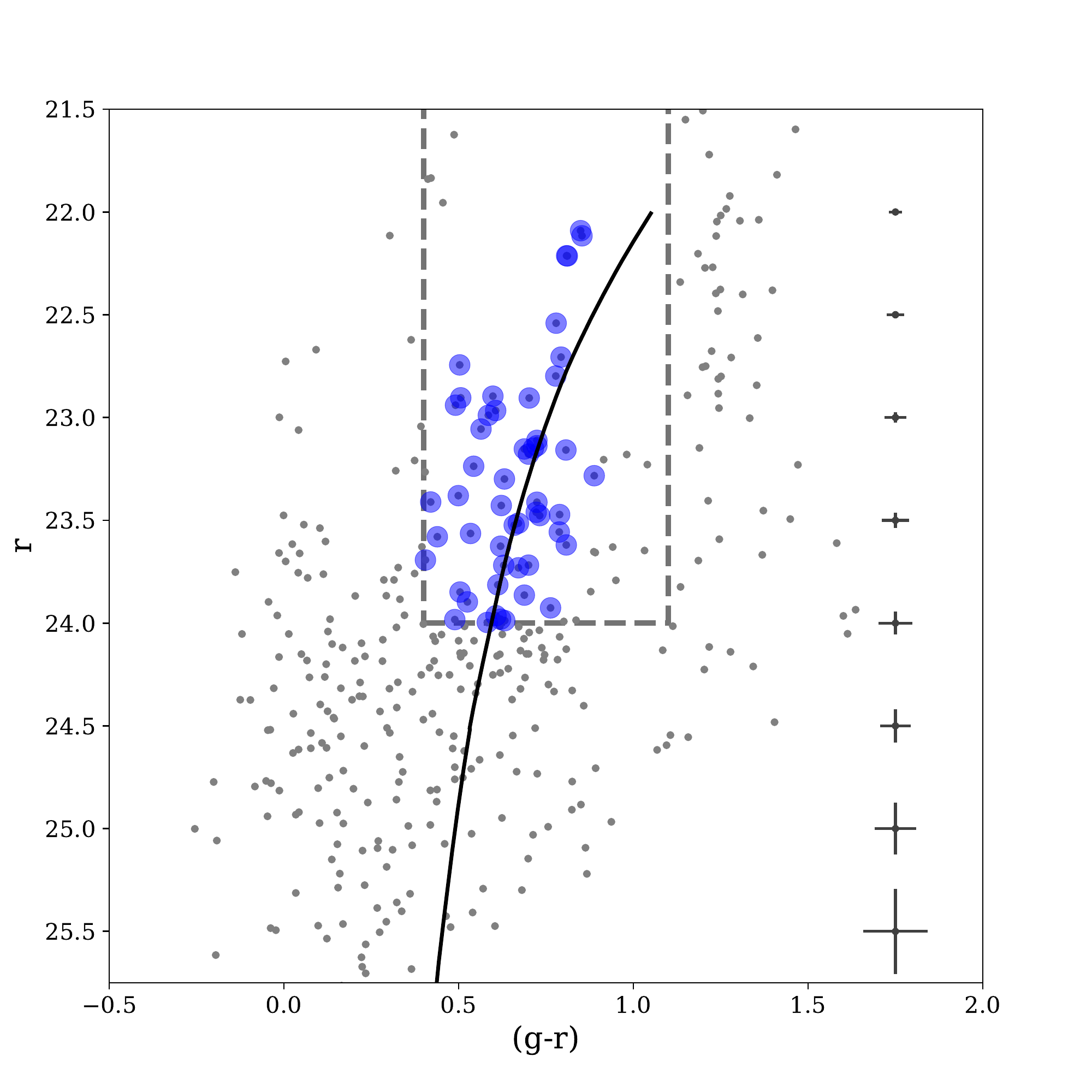}
	\includegraphics[width=\columnwidth]{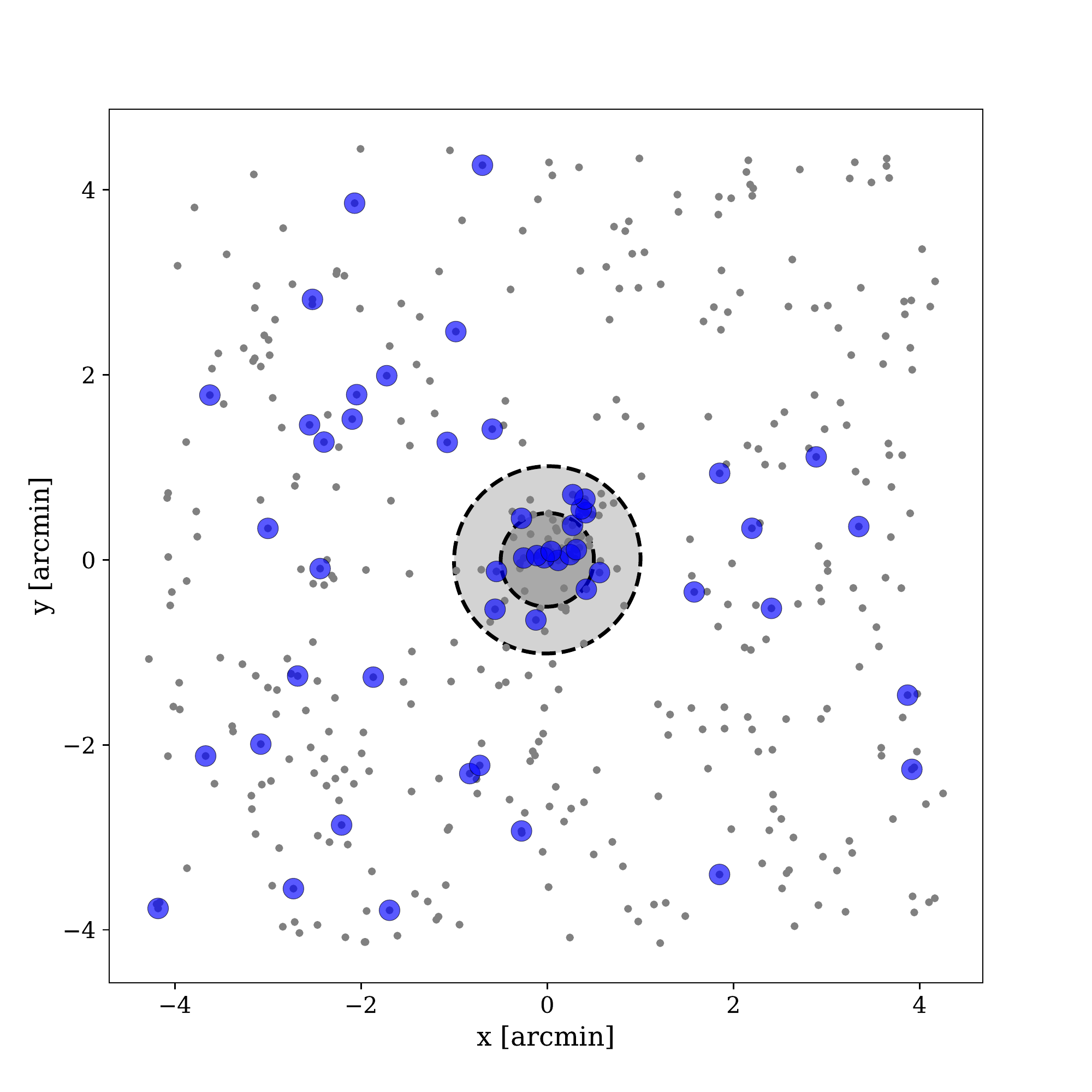}
      \caption{{\bf Left:} Colour magnitude diagram within ~1' of the centre of Pisces VII/Tri III. The grey data points indicate the complete data set observed. The blue data points are the sources deemed likely members of the candidate dwarf galaxy. The solid black line is an isochrone from the Dartmouth isochrone database with [Fe/H] = -2.0, dex [$\alpha$/Fe] = 0.4 dex and age = 12 Gyr. The grey dashed lines indicate the CMD selection box used. {\bf Right:} Spatial density plot of the observed data, with blue points showing the likely members of the dwarf galaxy which pass the criteria described in \S~\ref{sec:Identifying Probable Members}. The dashed circles shaded grey show $1\times$ and $2\times r_{\rm h}$.}
    \label{fig:CMD}
\end{figure*}

To determine the structural properties of the dwarf candidate, we identify stars that are most likely to be associated with the dwarf from the CMD. We select stars that lie on or near the RGB of the system. Three selection criteria were used to filter the foreground and background contaminants out of the data. First, the morphology of each datum was assessed and only stellar (point-like) sources were included. Second, a CMD selection box consisting of a color cut $(0.4 < g-r < 1.1)$, and limiting magnitude $(r < 24.0 )$ was introduced\footnote{We assume that this sample
is photometrically complete for this magnitude cut-off, that is $\sim$ 1.5 magnitudes brighter than the magnitude limit of our diagram.} . Finally, we use an old, metal-poor isochrone to trace the RGB (12 Gyrs, [$\alpha$/Fe]=0.4 dex, [Fe/H]=-2.0 dex taken from the Dartmouth isochrones, \citealt{dartmouth}), and measure the minimum distance, $d_{min}$, of each datum to the isochrone to calculate a probability of membership defined by: 

\begin{equation}
    P_{iso} = \exp{ \left[-\left(\frac{d_{min}^2 }{2\eta^2}\right)\right]}
    \label{eq:prob iso}
\end{equation}

where $\eta$ is a free parameter used to account for scatter about the isochrone. We set $\eta =$ 0.01 and use a low probability cut of $P_{iso} >$ 0.05 to isolate the most likely RGB candidates. This allows a broad selection of stars, without including too many obvious foreground contaminants by constraining the wider CMD selection box to the region around the isochrone. If a datum satisfied these criteria it is considered a probable member of the dwarf galaxy. A total of 52 probable member stars were identified within $\sim 1^\prime$ of the centre of the overdensity, as illustrated in Fig. \ref{fig:CMD}.


\subsection{Structural Properties}
\label{sec:Structural Properties} 

The structural properties were determined using the iterative Bayesian approach of Markov Chain Monte Carlo (MCMC) analysis and we use the \texttt{emcee} code developed by \citet{2013PASP..125..306F} to implement this. Our methodology followed the procedure outlined in \cite{martin16}, summarised below.

The radial density profile of the dwarf galaxy, $\rho_{\rm dwarf}(r)$, can be described by:

\begin{equation}
    \rho_{\rm dwarf}(r)=\frac{1.68^2}{2\pi r_{\rm h}^2(1-\epsilon)}N^*\exp{\left(\frac{-1.68r}{r_{\rm h}}\right)}
    \label{eq:density profile}
\end{equation}

where $\epsilon$ is the ellipticity, defined by the minor-to-major axis ratio, $\epsilon = 1- (b/a)$ and $r_{\rm h}$ is the half-light radius. N$^*$ is the number of likely member stars inside the CMD selection box associated to the dwarf, $r$ is the elliptical radius such that: 

\begin{equation}
    \begin{split}
        r= \Bigg( \Big(\frac{1}{1-\epsilon}((x-x_{0})\cos{\theta}-(y-y_{0})\sin{\theta})\Big)^2 \\
            +\Big((x-x_{0})\sin{\theta}-(y-y_{0})\cos{\theta}\Big)^2 \Bigg) ^{\frac{1}{2}}
    \label{eq:radial profile}
    \end{split}
\end{equation}
where $\theta$ is the position angle of the major axis, $x_{0}$ and $y_{0}$ are co-ordinates for the centre of the candidate dwarf galaxy and $x$ and $y$ are the coordinates on the plane tangent to the sky at the center of the DOLoRes field. The background contamination is considered constant and is determined by subtracting the contribution of the dwarf, which is calculated by performing a normalized  integration of the radial density profile from the total number of potential members identified, n. Hence, the background, $\Sigma_{b}$, can be described by: 

\begin{equation}
    \Sigma_{b} = \frac{\left(n - \int_{A}^{} \rho_{\rm dwarf} \,dA\right)}{\int_{A}^{}dA} .
    \label{eq:background}
\end{equation}

Equations \ref{eq:density profile}, \ref{eq:radial profile} and \ref{eq:background} combine together to give the following likelihood function, used in the MCMC analysis:

\begin{equation}
    \rho_{\rm model}(r) = \rho_{\rm dwarf}(r) + \Sigma_{b} .
    \label{eq:total profile}
\end{equation}

\begin{figure}
	\includegraphics[width=\columnwidth]{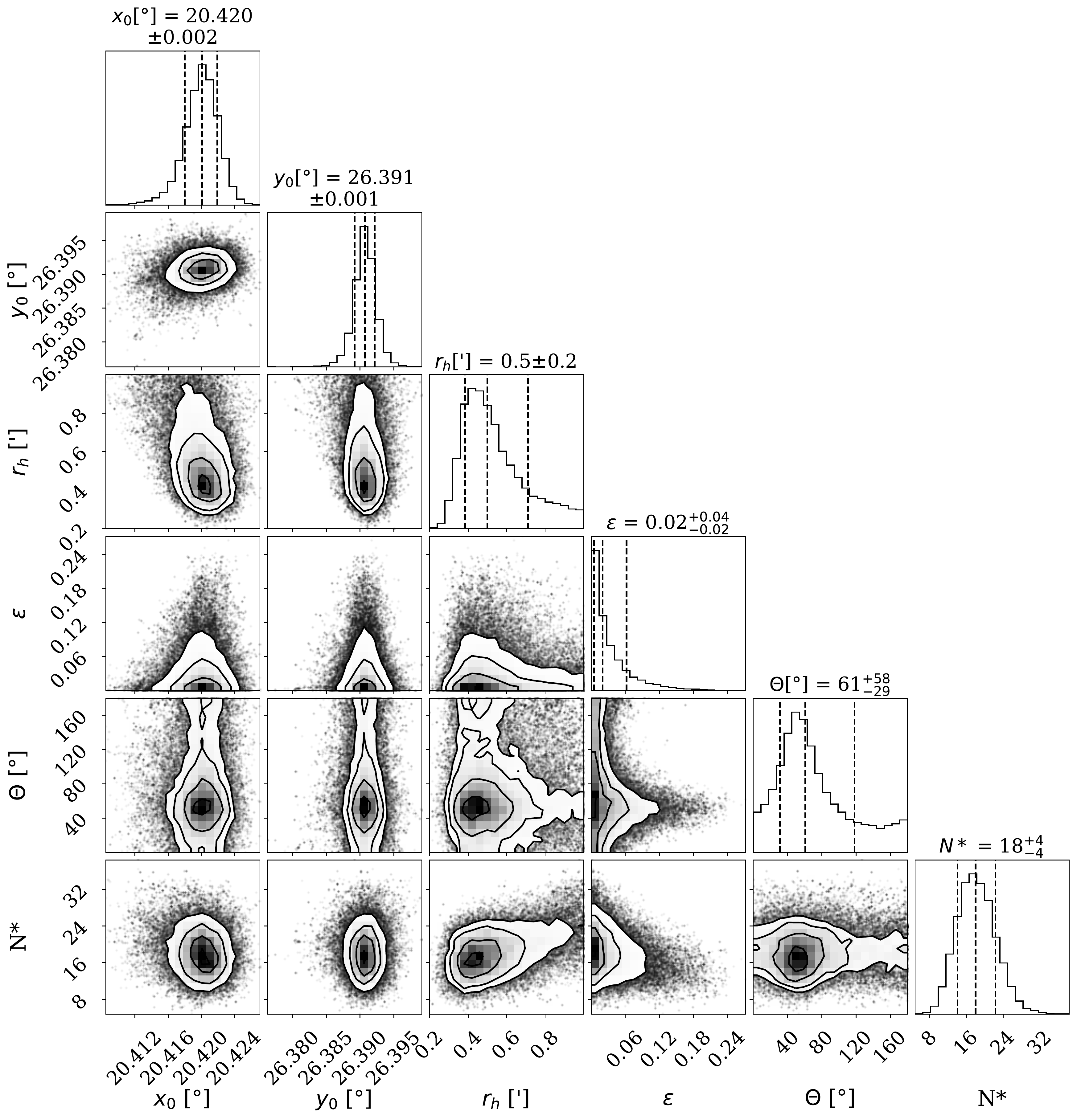}
    \caption{2D and marginalized PDFs for the central coordinates of the system, $x_{0}$ and $y_{0}$, the half-light radius, $r_h$, the ellipticity, $\epsilon$, the position angle of the major axis, $\theta$ and the number of stars belonging to the system from the CMD selection criteria, N$^*$ . The dashed lines represent the mean value and 1$\sigma$ uncertainties.}
    \label{fig:corner plot}
\end{figure}

 Uniform flat priors were used to constrain the parameter space to physical solutions but were kept broad to ensure the analysis wasn't over-constrained, see Table \ref{tab:priors}. The \texttt{emcee} routine used 100 walkers, over a total of 10,000 iterations with a burn in stage of 8,500. Fig. \ref{fig:corner plot} is the resulting corner plot from the MCMC analysis and the derived structural parameters are summarised in Table \ref{tab:structural params}. 

\begin{table}
	\centering
	\caption{The priors used in the \texttt{emcee} routine.}
	\label{tab:priors}
	\begin{tabular}{ll}
		\hline
		MCMC Parameter & Value\\
		\hline
		$x_{0}$ (radians) & 0.3 $\leqslant x_{0} \leqslant 0.4$\\
		$y_{0}$ (radians) & $0.45 \leqslant y_{0} \leqslant 0.5$\\
		$r_{\rm h}$ (arcmin) & $0 \leqslant r_{h} < 1$ \\
		$\epsilon$ & $0 \leqslant \epsilon \leqslant 1$\\
		$\theta$ (radians) & $0 \leqslant \theta \leqslant \pi$\\
		N$^{*}$ & 0 < N$^{*}$ $\leqslant$ n\\
		\hline
	\end{tabular}
\end{table}


\begin{table}
	\centering
	\caption{The final structural and photometric properties for the dwarf.}
	\label{tab:structural params}
	\begin{tabular}{ll}
		\hline
		Property & Value\\
		\hline
		RA & $1^{h}$ $21^{m}$ $40.5^{s}$ $\pm$ ${0.8^{s}}$\\
		Dec & 26\textdegree $ 23^{'}$ $ 24^{"}$ $\pm$ ${4^{"}}$\\
		$r_{\rm h}$ (arcmin) & 0.5 $\pm$ {0.2}\\
		$D$ (kpc) & $1000^{+300}_{-200}$\\
		$r_{\rm h}$ (pc) & $131\pm61$\\ 
		M$_V$ & $-6.8\pm0.2$\\
		$\mu_0$ (mag arcsec$^{-2}$) & $27.7\pm0.3$\\
		$L$  ($L_\odot$) & $3.7^{+0.8}_{-0.6}\times 10^4$\\
		$\epsilon$ & 0.02 $^{{+{0.04}}}_{{-{0.02}}}$\\
		$\theta$ (\textdegree) & 61 $^{{+{58}}}_{{-{29}}}$\\
		N$^*$ & 18 $^{{+{4}}}_{{-{4}}}$\\
		\hline
	\end{tabular}
\end{table}


\begin{figure}
	\includegraphics[width=\columnwidth]{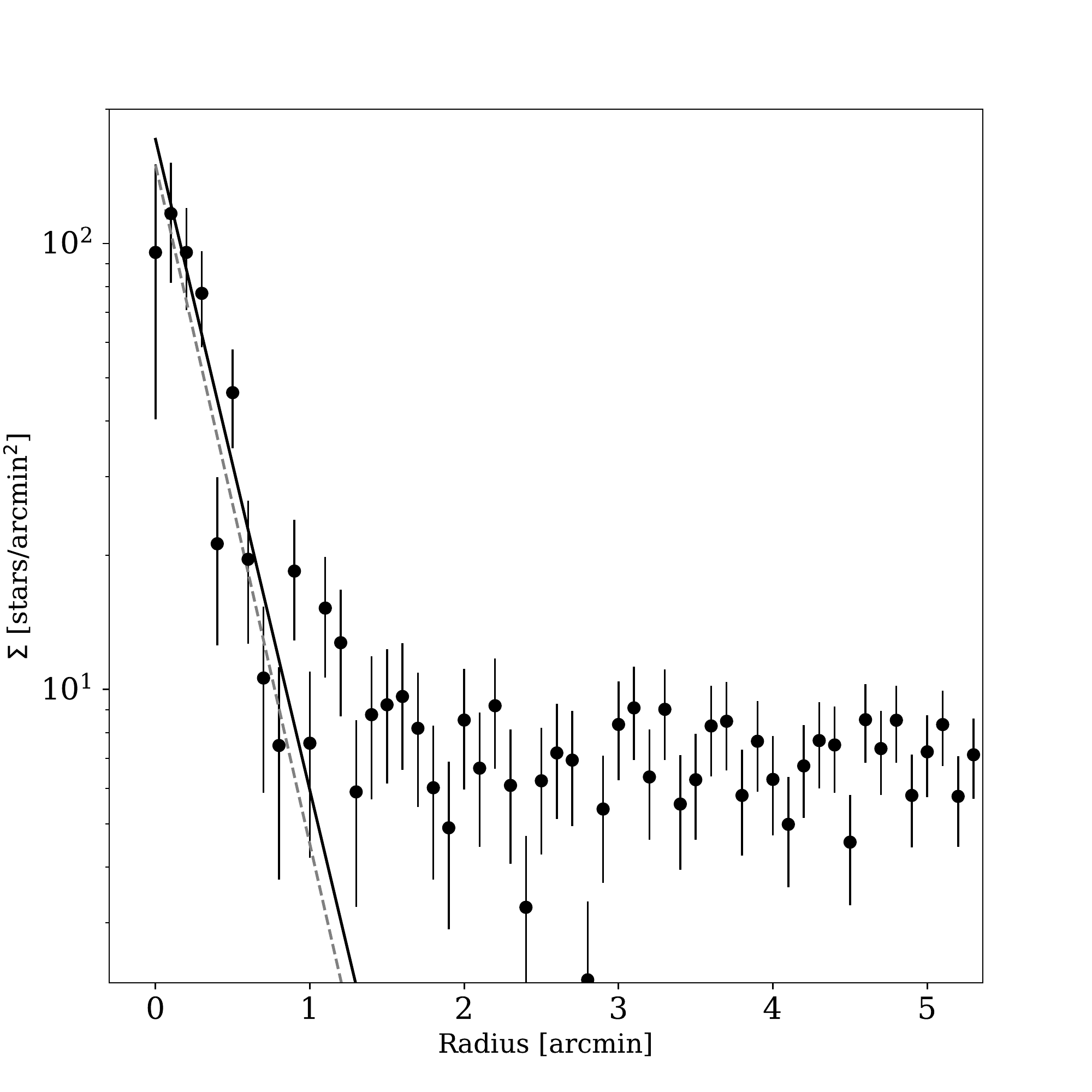}
    \caption{Stellar radial density profile with the observed data binned in elliptical annuli with the favoured structural properties of ellipticity, the position angle, the structural centre and the number of stars. The solid black fit uses the MCMC values and the dashed grey line is the chi-squared fit, which yields a value for $r_{\rm h} = 0.48^\prime$. }
    \label{fig:radialprofile}
\end{figure}

Next we plot a stellar radial density profile using the parameters that were deduced from the MCMC analysis, which we show in Fig.~\ref{fig:radialprofile}. The error bars on the data points represent the Poisson uncertainties for each point. We overplot the exponential profile deduced using MCMC, which agrees very well with the observed profile. In addition we perform a basic chi-squared fit of the exponential profile shown in Equation~\ref{eq:density profile} to this binned data. We recover a very similar value for $r_{\rm h}=0.48^\prime$, completely consistent with the MCMC approach.

\subsection{The distance to the dwarf}
\label{sec:distance}

\begin{figure}
	\includegraphics[width=\columnwidth]{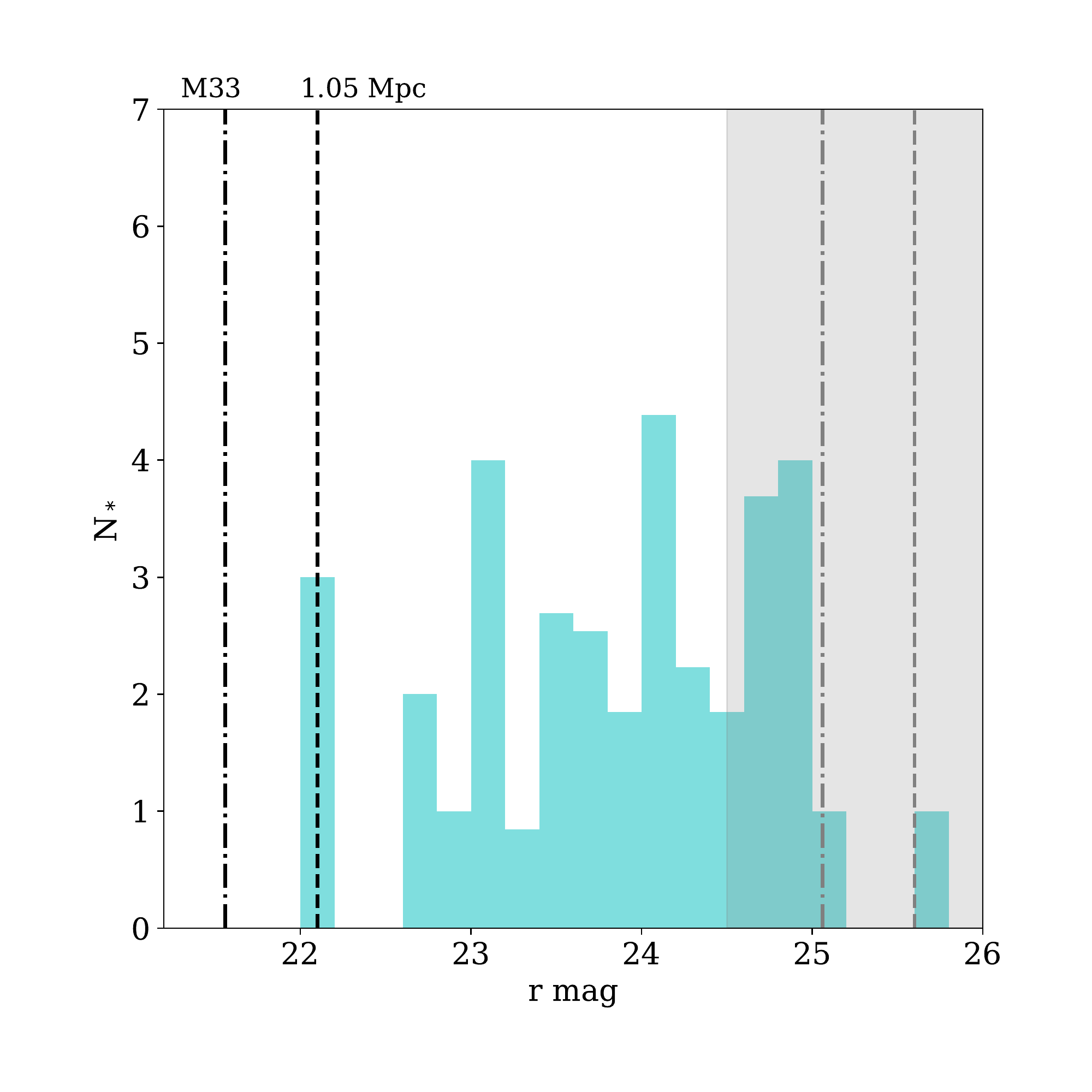}
    \caption{A background corrected luminosity function for all stars within $2\times r_{\rm h}$ of the galaxy centre. The gray shaded region shows the point at which our completeness drops below 90\%. Using a Sobel edge detection filter, we find a peak in the luminosity function at $r=22.1$, which is indicated as the dashed black line. The location of the TRGB of M33 is shown as a dot-dashed line. The grey lines show the predicted locations for the HB in the LF for these three distances.}
    \label{fig:LF}
\end{figure}

To determine the distance to Pisces VII/Tri III, we use the Tip of the Red Giant Branch (TRGB) method. We construct a luminosity function for our dwarf galaxy (see Fig.~\ref{fig:LF}), using all sources within $2\times r_{\rm h}$ of the galaxy's centroid and colours of $0.3<g-r<1.1$. This produces a reasonably tight selection of RGB stars, and minimises foreground contamination. From this, we subtract an area-normalised background luminosity function constructed from sources with the same colour cut, between 2.5-3.5 arcmin from our source. The gray shaded area in Fig.~\ref{fig:LF} shows the region below which our photometry is $<90\%$ complete.

\begin{figure*}
	\includegraphics[width=\textwidth]{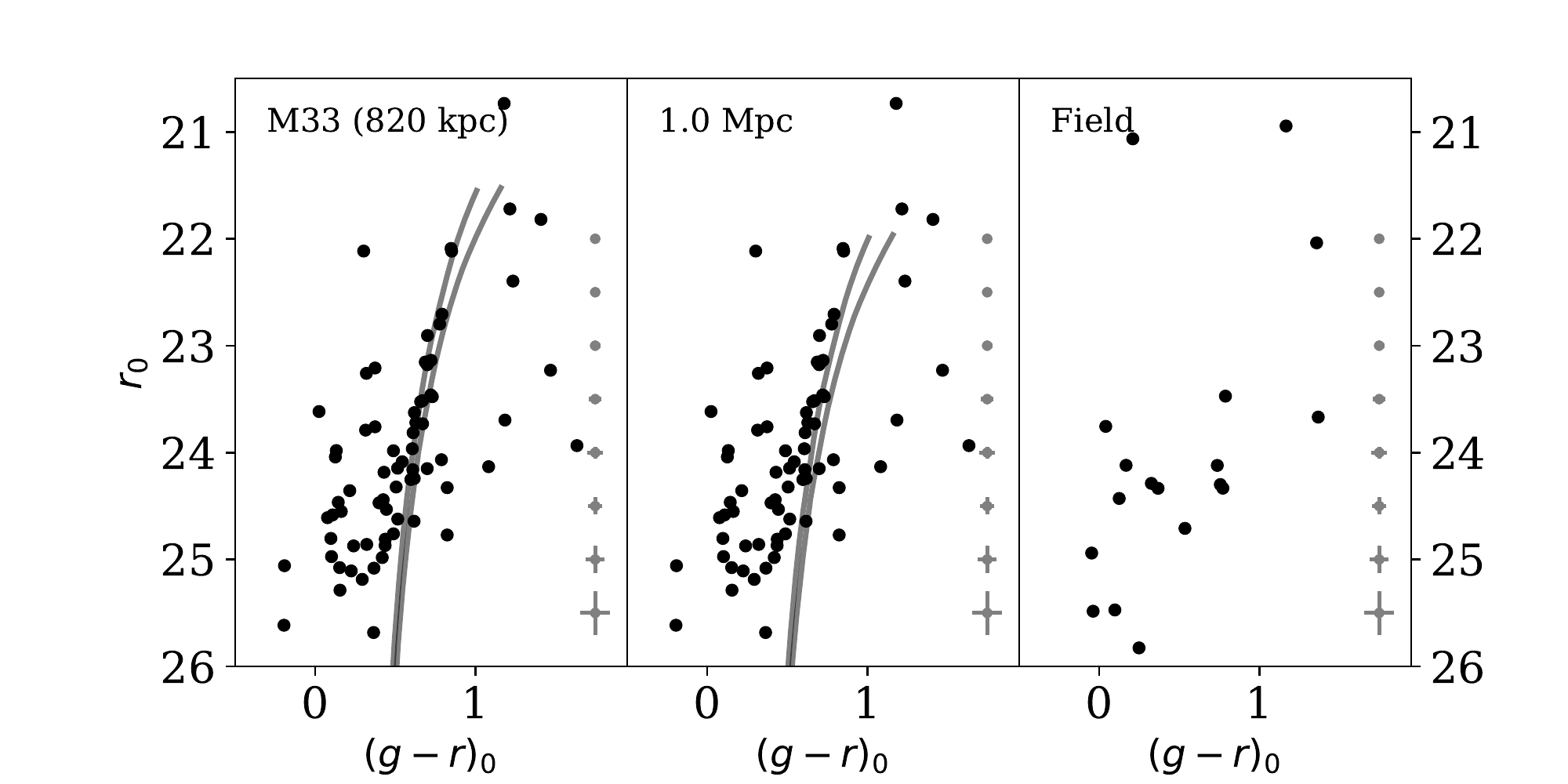}
    \caption{The first 2 panels show the CMD for all sources within $2\times r_{\rm h}$ of the centre of the dwarf. In each panel, \citet{dartmouth} isochrones with an age of 12 Gyr, $[\alpha/{\rm Fe}]$ = +0.4 and metallicities of [Fe/H]$ =-2.0\,{\rm and}\,-2.5$ are overlaid for two different distances: The distance to M33 (820 kpc), and that from our TRGB analysis (1.0~Mpc). The final panel shows a control field located away from the centre of the dwarf of equal area. Using the TRGB estimate, the isochrones imply the system is very metal poor, whereas the M33 distance gives a good match to the colours of the RGB stars.}
    \label{fig:dCMD}
\end{figure*}

The TRGB in the $r-$band should have an absolute magnitude of $M_{r, {\rm TRGB}}=-3.01\pm0.1$ \citep{sand14}. To begin with, we apply a Sobel edge detection filter to the luminosity function \citep{lee93}. This has been a historically popular way of finding the sharp edge of the TRGB, but it doesn't consider any uncertainties in the data. From this, we a value of possible values for the TRGB at $r=22.1$, which is marked as the black dashed line in Fig.~\ref{fig:LF}. This corresponds to physical distance estimate of 1.05~Mpc. Owing to the paucity of stars within our dataset, this method has merely located the 3 brightest stars in the dataset. The dot-dashed line represents the TRGB location for an object at the distance of M33 (820~kpc, \citealt{conn12}). In Fig.~\ref{fig:dCMD}, we show the CMD for all stars witin $2\times r_{\rm h}$ of the dwarf galaxy with isochrones of varying metallicity overlaid at each of these proposed distances \citep{dartmouth}. For a distance of 820~kpc, the dwarf appears metal poor, with [Fe/H]$\sim -2.0$.

As a second step, we could also look for a peak in the luminosity function at the location of the horizontal branch (HB). The HB is a secure distance estimator and it is typically better populated than the RGB, allowing a better constraint. In Fig.~\ref{fig:LF}, we also show the predicted location of the HB assuming the Sobel distance estimate, as well as the HB for an object at the distance of M33. However, as our photometry is incomplete at these magnitudes, we cannot locate this feature. 


A more rigorous way of searching for the TRGB is to use Bayesian inference (e.g. \citealt{conn12}). Here, we follow the method outlined by \citet{tollerud16a}, who model the RGB and background population simultaneously as a broken power law in luminosity, with the break located at the TRGB. The luminosity function is also convolved with the photometric uncertainties, providing an advantage over the Sobel method above. We adapt their public TRGB fitting code (see \citealt{tollerud16b}) for our dataset, and summarise their method briefly below. The assumed luminosity function, $\phi_{\rm RGB}$ is defined as

\begin{eqnarray}
\label{eq:rgblf}
&\phi_{\rm TRGB}(m,m_{\rm trgb},\alpha,\beta,f)\nonumber\\
= & \begin{cases}
      \frac{\alpha e^{\alpha(m-m_{\rm trgb})}}{e^{\alpha(m_2-m_{\rm trgb}}-e^{\alpha(m_1-m_{\rm trgb}}} & \text{if}\ m> m_{\rm trgb} \\
      f \frac{\beta e^{\beta(m-m_{\rm trgb})}}{e^{\beta(m_2-m_{\rm trgb}}-e^{\beta(m_1-m_{\rm trgb}}}& \text{if}\ m\le m_{\rm trgb}
    \end{cases}
\end{eqnarray}

\noindent where $m$ is the stellar magnitude, $m_{\rm trgb}$ is the TRGB magnitude, $\alpha$ and $\beta$ describe the power law slopes of the RGB and background respectively, and $f$ gives the fraction of stars in the background population (such that the fraction on the RGB is $1-f$). We then write the per star likelihood as

\begin{eqnarray}
\label{eq:likelihood}
\mathcal{L}(m_i \mid m_{\rm trgb},  \alpha, \beta, f)\nonumber\\
= N\int^{m_2}_{m_1}\phi_{\rm RGB}(m_{\rm trgb}, \alpha, \beta, f) \\
\times \mathcal{N}\left[m_i- m, \sigma(m)\right]{\rm d}m\nonumber
\end{eqnarray}

\noindent where $m_i$ are the extinction-corrected magnitudes for each star, $m_1$ and $m_2$ are the bright and faint limits of the data, $N$ is a normalization constant (determined by numerical integration), $\mathcal{N}$ is the standard Gaussian distribution and $\sigma(m)$ are the per-star uncertainties.

Our primary interest is the value of $m_{\rm trgb}$, so we use Bayesian inference with this model to determine posterior probabilities for this using Equation~\ref{eq:likelihood} as our likelihood function. We implement uniform priors on $f$ of U(0, 1), and for $\alpha$ and $\beta$ we use the same priors as \citet{conn12} of U(0, 2). For $m_{\rm trgb}$ we adopt a more informative prior based on the results of our Sobel filter method above, setting this to U(20, 23.1). For $m_i$, we use our $r-$band magnitudes and use the same colour selection function as for the luminosity function above ( $0.5<g-r<1.1$), and only include stars with $r<24$ to ensure we are using a complete sample. We then numerically integrate Equation~\ref{eq:rgblf} over a grid of $m$ using the trapezoid rule for a given set of parameters, and take the total likelihood as the sum of the logarithm of this over all stars.

\begin{figure*}
	\includegraphics[width=\columnwidth]{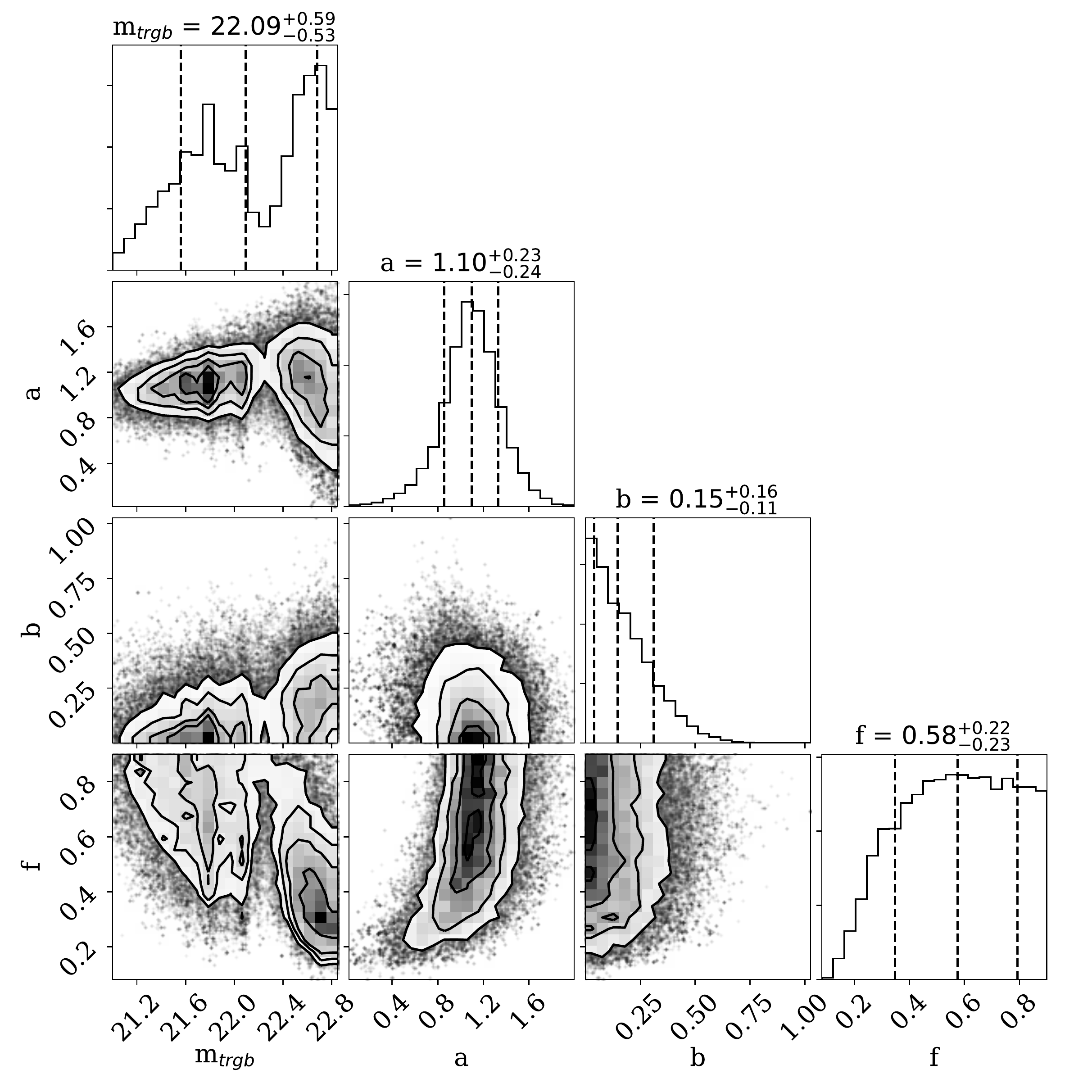}
	\includegraphics[width=\columnwidth]{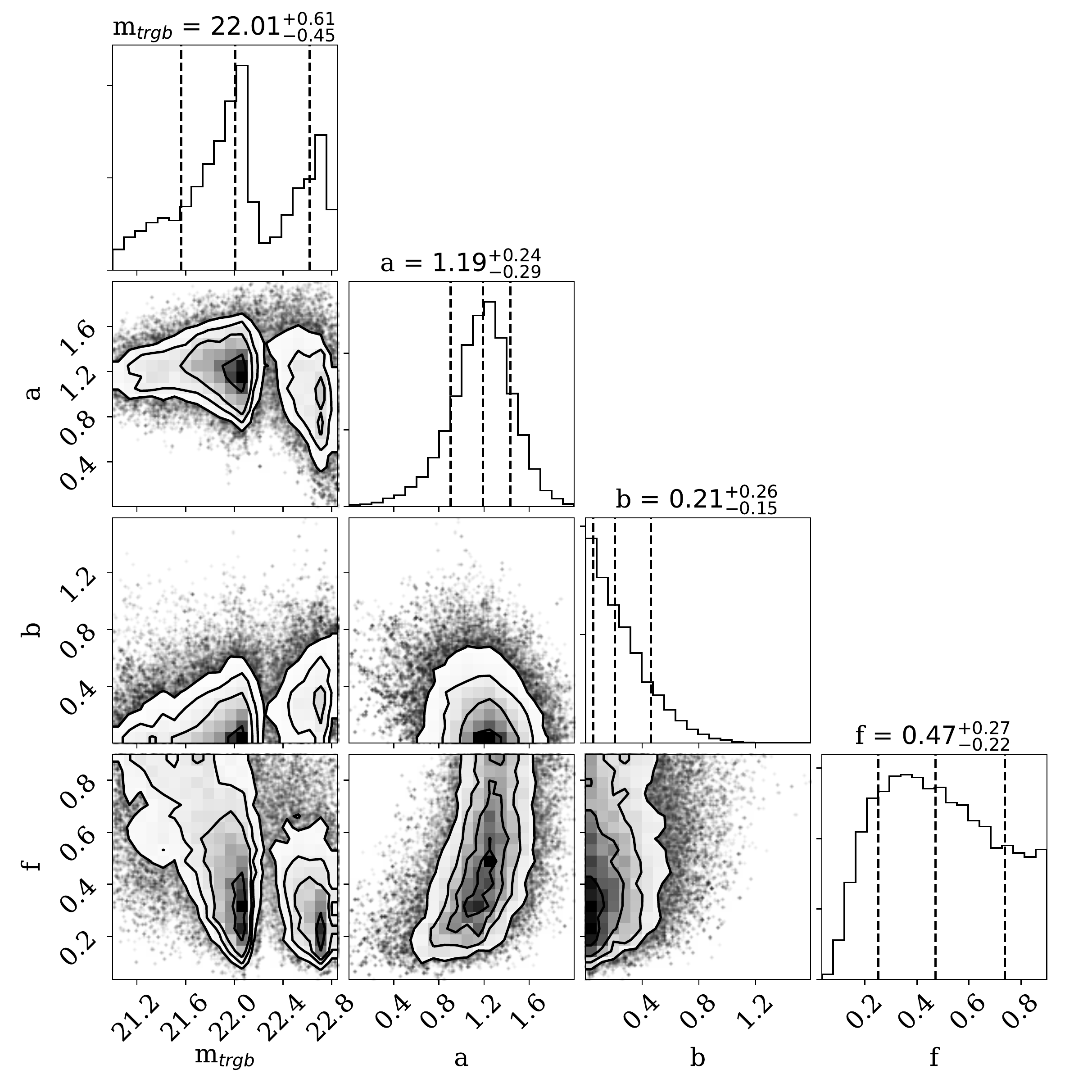}
    \caption{{\bf Left:} Posterior distribution for the MCMC TRGB finding algorithm using our standard colour cut. The key parameter of interest is $m_{\rm trgb}$. For this colour cut, the distribution shows 3 peaks, and is poorly constrained, resulting in a distant estimate between 0.9--1.4 Mpc. {\bf Right:} The same, but for a marginally tighter colour cut ($0.5<g-r<1.1$). Here, $m_{\rm trgb}=22.0^{+0.6}_{-0.5}$, favouring an isolated distance of $D=1.0^{+0.3}_{-0.2}$~Mpc. The posterior results are very sensitive to assumptions of priors and colour cuts, owing to the paucity of stars (dwarf and background model) in the region. }
    \label{fig:dcorner}
\end{figure*}

We again use {\sc emcee} to sample the posterior distribution for the model parameters for this broken power-law distribution. The results of this are shown in the left panel Fig.~\ref{fig:dcorner}. The results are inconclusive, and highly dependent on both the priors and the colour and distance selection. Using purely the colour cut, and the whole field of view, we find a triple-peaked distribution, with higher posterior values for $m_{\rm trgb}\sim21.8, 22.1$ and $22.7$ (corresponding to distances of $\sim0.9-1.4$~Mpc). If we narrow the colour cut slightly to exclude the bluer stars which may or may not be members ($0.5<g-r<1.1$), we instead recover the Sobel value for the TRGB of of $m_{\rm trgb}=22.0^{+0.6}_{-0.5}$ (right panel Fig.~\ref{fig:dcorner}), corresponding to a distance of $D=1.0^{+0.3}_{-0.2}$~Mpc. If we only include stars within the central two half-light radii (to exclude background contaminants), we measure $m_{\rm trgb}=21.9\pm0.5$ ($D=960^{+200}_{-190}$~kpc).

This sensitivity to our selection criteria is driven by the scant number of stars both within the dwarf galaxy itself, and in the surrounding field. The model is highly sensitive to individual stars, and the uncertainties in all cases are likely underestimated. The stellar populations are scarce, making it difficult to distinguish between foreground and dwarf stars (as shown by the $f$ parameter in both corner plots). Overall, this approach appears to favour an isolated distance for Pisces VII/Tri III. To confirm this, we require deeper data which would allow us to resolve the better populated horizontal branch or main sequence turn-off. For the rest of the paper, we assume a distance for the dwarf of $D=1.0^{+0.3}_{-0.2}$~Mpc, and use these to calculate a physical size and luminosity range for this candidate.

\subsection{Luminosity}
\label{sec:luminosity}


 The luminosity of Pisces VII /Tri III was calculated using a methodology similar to the approach outlined in \cite{martin16}. A probability density function (PDF) was made using a theoretical luminosity function, describing the number of expected star counts per narrow magnitude bin for a stellar population with an age of 12 Gyr, $[\alpha/{\rm Fe}]$ = +0.4 dex and metallicity of [Fe/H] $= -2.0$ dex. This PDF was then used to weight random sampling in magnitude space. Stars which are selected from the isochrone which have a magnitude above the magnitude limit on the CMD used in our analysis (r < 24.0) are flagged. The magnitude is then converted into the corresponding luminosity. When converting into a luminosity the uncertainty in the distance was considered by sampling the distance from a Gaussian distribution centered around the results from the distance modulus with a $\sigma$ corresponding to the uncertainty of the distance modulus. The process of random sampling was repeated until a total of N* flagged stars was obtained, matching our dataset. Once this limit was reached the luminosity of all sampled stars was summed, including those not flagged, to give a final luminosity value. The whole process was repeated 1000 times and an average luminosity was obtained. Assuming the isolated distance, the resulting magnitudes were $M_{r}=-6.5\pm{0.3}$ and $M_g=-6.8\pm{0.2}$. The uncertainty comes from the standard deviation of the 1000 iterations, wherein uncertainty is introduced in the random sampling of the distance Gaussian indicative of the distance error.  We then convert this into an absolute magnitude in the $V$-band using the colour transforms of \citet{jordi06}, giving $M_V=-6.8\pm{0.2}$. Assuming Pisces VII/Tri III is instead an M33 satellite, the resulting g and r band magnitudes are $M_r=-6.1\pm{0.2}$ and $M_g=-6.3\pm{0.3}$, giving $M_V=-6.1\pm{0.2}$.

From these values, we find the total luminosity of the dwarf galaxy to be $L=2.1\pm0.4\times10^4\,{\rm L_\odot}$ and $L=3.7^{+0.8}_{-0.6}\times10^4\,{\rm L_\odot}$ for the M33 and isolated distances respectively. We also calculate the distance independent central surface brightness of the dwarf galaxy $\mu_0$, and find a value of $\mu_0 = 27.6\pm0.2$ mag arcsec\ensuremath{^{\mathrm{-2}}} .


\begin{figure*}
	\includegraphics[width=\columnwidth]{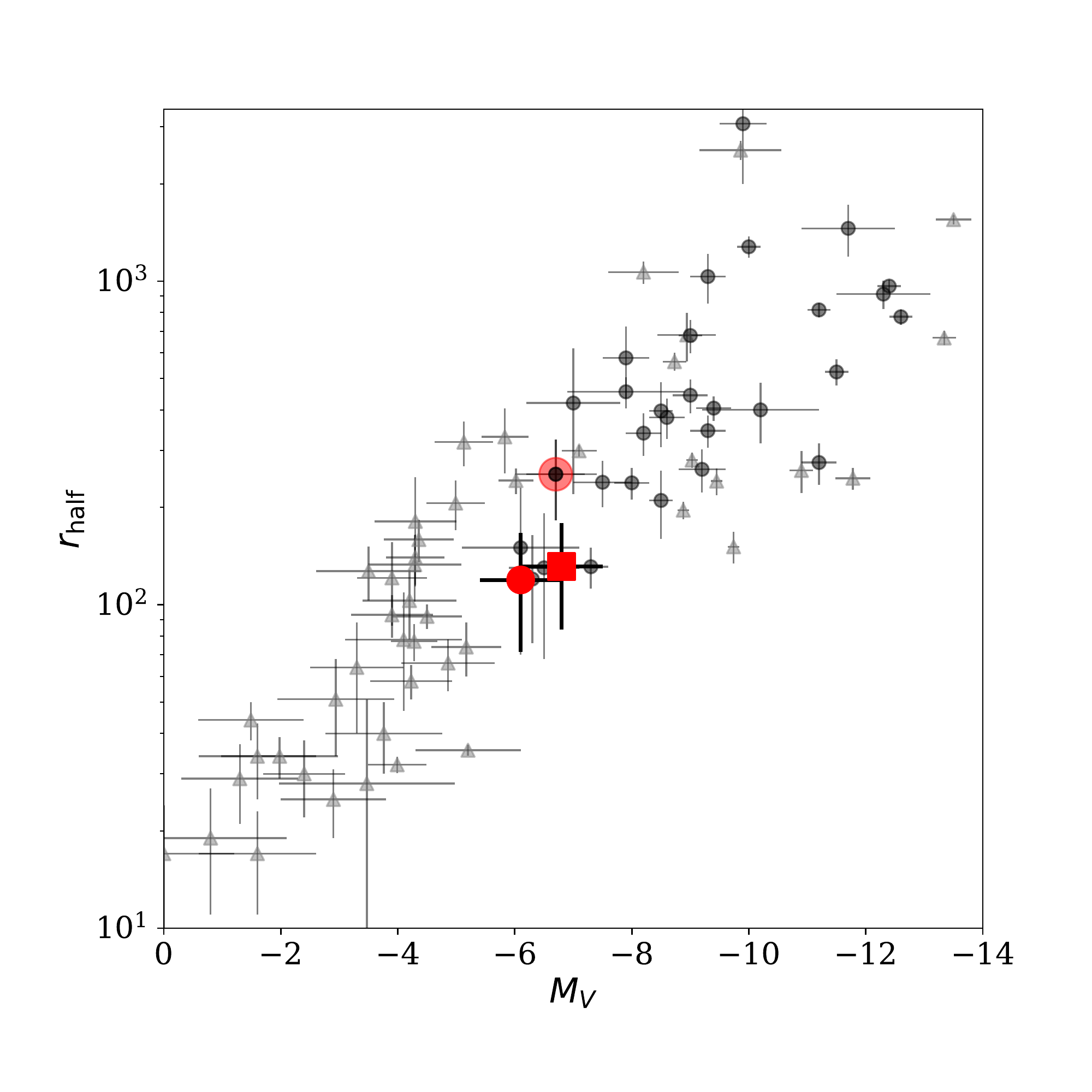}
	\includegraphics[width=\columnwidth]{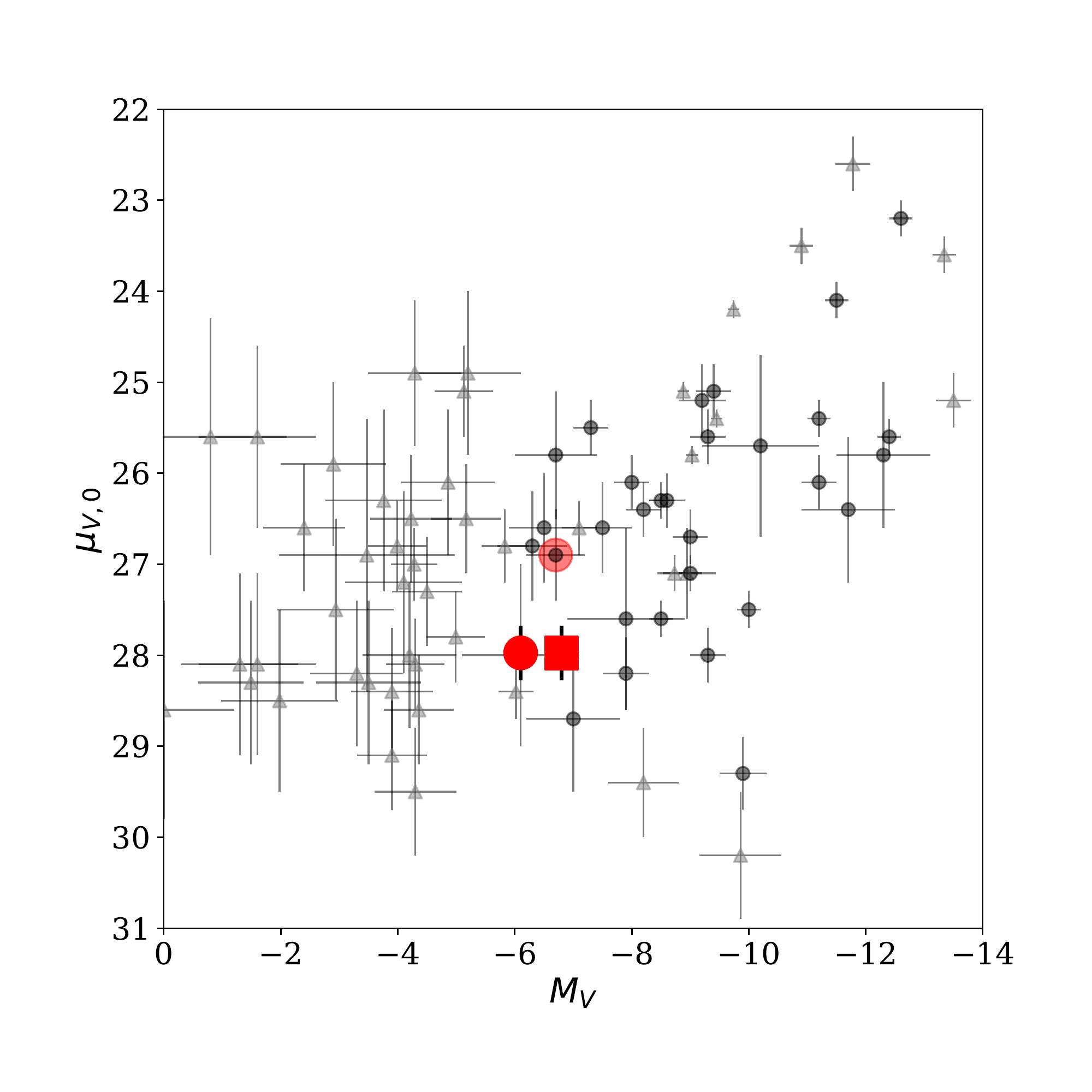}
    \caption{{\bf Left:} $M_V$ vs $r_{\rm h}$ for dwarf spheroidal galaxies of the Milky Way (light grey triangles) and M31 (dark gray circles). Our new candidate is highlighted as a large red circle (for an M33 distance) and a red square (for the isolated distance), and fits comfortably within the size-luminosity relation for Local Group dwarfs. The point with the red shading is the only other likely satellite of M33, Andromeda XXII/Triangulum I. {\bf Right: } $M_V$ vs central surface brightness, $\mu_{V,0}$, for Local Group dwarfs.}
    \label{fig:summary}
\end{figure*}

\section{DISCUSSION AND CONCLUSIONS}
\begin{figure}
	\includegraphics[width=\columnwidth]{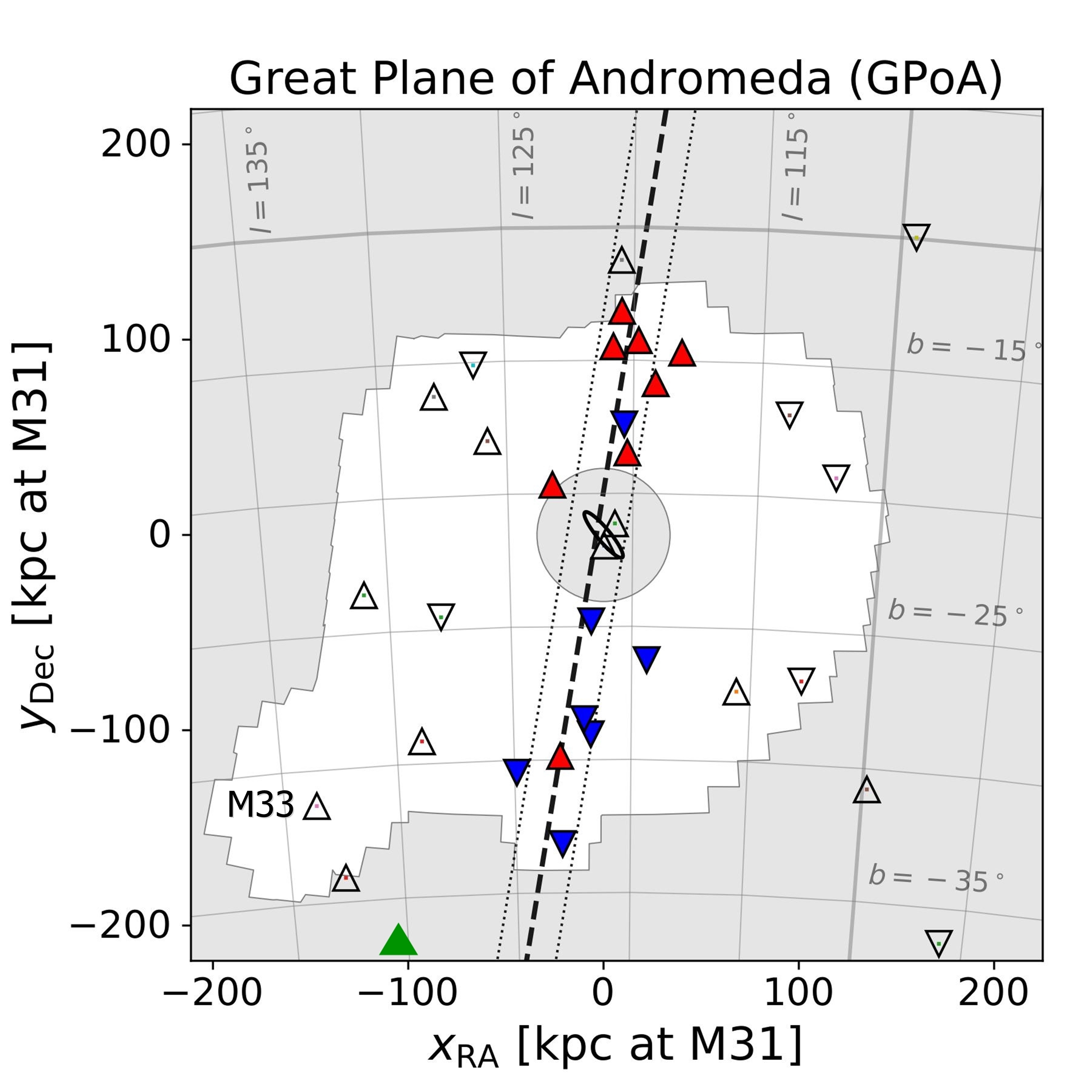}
    \caption{Position of Pisces VI/Tri III ({\it green triangle}) with respect to M31 (black ellipse) and M33 and the GPoS. The dashed and dotted lines indicate the orientation and width of the best-fit GPoS respectively. The white footprint corresponds to the region explored by the PAndAS survey. Satellites with measured radial velocities are showed as color-coded triangles, according to whether they are approaching (blue) or receding (red) relatively to M31. Adapted from Fig. 1 in Pawlowski (2018).}
    \label{fig:SP}
\end{figure}

We report the discovery of Pisces VII/Tri III, a new dwarf galaxy in the surroundings of M33 by visual inspection of the public available image of the DESI LIS outside the PAndAs survey footprint. Using deeper follow-up imaging from DOLoRes@TNG we attempt to constrain a distance for this dwarf. The paucity of stars makes this extremely challenging, but based on both a Sobel and MCMC approach, the likely distance $D=1.0^{+0.3}_{-0.2}\,{\rm Mpc}$, making it either a satellite of M33, or an isolated ultra-faint dwarf (UFD) in the field. It is faint and compact with $M_V=-6.1\pm0.2$ for the M33 distance or $M_V=-6.8\pm0.2$ for the isolated case. It would then have a half-light radius of  $r_{\rm h}=119\pm48\,{\rm pc}$ for an M33 distance or $r_{\rm h}=131\pm61\,{\rm pc}$ for the isolated case, comparable to UFDs of the Milky Way (see Fig.~\ref{fig:summary}). This luminosity would make it one of the faintest dwarf galaxies detected beyond the Milky Way, and potentially the faintest detected field galaxy to-date.

Armed with the structural and photometric properties of our dwarf, we can place it in context with the other dwarf galaxies of the Local Group. In Fig.~\ref{fig:summary}, we show the relationship between luminosity and half-light radius (left) and surface brightness (right). The light grey triangles show Milky Way satellites, while dark grey circles show M31 satellites. These values are compiled from a number of sources \citep{mcconnachie12, koposov15,bechtol15,drlica15,martin16,weisz19}. Our candidate is highlighted as a large red circle and square for the M33 and isolated distances respectively. It is only the second potential satellite galaxy of M33 discovered. The other, Andromeda XXII, is highlighted with red shading\footnote{Laevens~2/Tri II is a Milky Way dwarf galaxy situated at 30 kpc (\citealt{2015ApJ...802L..18L,kirby15,martin16b,kirby17}).}.

Our derived properties for the new dwarf are perfectly consistent with other ultra-faint dwarf galaxies. If this galaxy was confirmed as a satellite of M33, it would alleviate the current tension between the observed and predicted number of satellites around M33 \citep{patel18}. Given its low luminosity, it also suggests it may be the tip of the iceberg in terms of finding more ultra-faint dwarfs in the M31-M33 system.

The new census of dwarf galaxies and the homogeneous distance measurements for all the known M31 satellites from
the PAndAS survey suggest the possible existence of a coherent flattened galaxy plane of 15 satellite galaxies in that survey
volume \citep{ibata13}. This Great Plane of Andromeda (GPoA, see Fig~\ref{fig:SP}) is almost edge-on
orientated and extends more than 400 kpc from the center of M31. Interestingly, this plane seems to be aligned with the Giant Stellar Stream in the M31 halo and, in contrast with a similar satellite plane found in the Milky Way \citep{pawlowski12}, it is not perpendicular to its galactic disk but inclined $\sim$ 50 degrees.  The existence of these kinematically correlated satellite planes in the Local Group spirals are  very rare in cosmological simulations based on the $\Lambda$CDM paradigm (e.g. \citealt{pawlowski18}). Some solutions proposed for this possible
small-scale problem for the $\Lambda$CDM theory are the accretion of dwarf galaxies along filaments of the cosmic web (e.g. \citealt{buck15}), infall of satellites in groups (e.g. \citealt{samuel20}) or a possible tidal dwarf galaxy origin for some of the satellites of Andromeda \citep{hammer18}.

The available DESI LIS imaging data outside the PAndAs footprint also allows us to explore the existence of additional members of this possible GPoA at larger projected distances from M31. Fig. \ref{fig:SP} shows the position of the Pisces VII/Tri III with respect to M33 and the GPoA (Pawlowski 2018). It clearly lies off the plane, and in projection seems likely associated to the outer M33 halo.

The discovery of Pisces VII/Tri III by visual inspection of a limited area around M33 using the DESI LIS deep imaging suggests that there is still room for discovery of low surface brightness dwarf galaxies lurking in the outskirts of Andromeda, beyond the bounds of the PAndAS survey. To confirm whether it is a bona-fide satellite of M33, a precision distance should be measured using deep imaging. Follow-up studies using the Hubble Space Telescope would allow the HB and main sequence of the dwarf to be resolved, allowing us to distinguish between the distances measured from the scarce RGB stars. In addition, spectroscopy of its brightest member stars would allow us to determine whether it is dynamically bound to M33. Given the faint nature of the stars, an 8--10~m class telescope, such as Keck, would be required. 

\section*{Acknowledgements}

We thank the TNG director, Dr Ennio Poretti, for the telescope time  he kindly granted us to perform the photometric follow-up of this dwarf galaxy. DMD acknowledges financial support from the Talentia Senior Program (through the incentive ASE-136) from Secretar\'\i a General de  Universidades, Investigaci\'{o}n y Tecnolog\'\i a, de la Junta de Andaluc\'\i a. DMD and EJA acknowledge funding from the State Agency for Research of the Spanish MCIU through the ``Center of Excellence Severo Ochoa" award to the Instituto de Astrof{\'i}sica de Andaluc{\'i}a (SEV-2017-0709). This publication is based on observations made on the island of La
Palma with the Italian Telescopio Nazionale {\it Galileo}, which is
operated by the Fundaci\'on Galileo Galilei-INAF (Istituto Nazionale
di Astrofisica) and is located in the Spanish Observatorio of the
Roque de Los Muchachos of the Instituto de Astrof\'isica de Canarias. 

This project used public archival data from the {\it DESI Legacy Imaging Surveys} (DESI LIS). The Legacy Surveys consist of three individual and complementary projects: the Dark Energy Camera Legacy Survey (DECaLS; Proposal ID 2014B-0404; PIs: David Schlegel and Arjun Dey), the Beijing-Arizona Sky Survey (BASS; NOAO Prop. ID 2015A-0801; PIs: Zhou Xu and Xiaohui Fan), and the Mayall z-band Legacy Survey (MzLS; Prop. ID 2016A-0453; PI: Arjun Dey). DECaLS, BASS and MzLS together include data obtained, respectively, at the Blanco telescope, Cerro Tololo Inter-American Observatory, NSF’s NOIRLab; the Bok telescope, Steward Observatory, University of Arizona; and the Mayall telescope, Kitt Peak National Observatory, NOIRLab. The Legacy Surveys project is honored to be permitted to conduct astronomical research on Iolkam Du’ag (Kitt Peak), a mountain with particular significance to the Tohono O’odham Nation. NOIRLab is operated by the Association of Universities for Research in Astronomy (AURA) under a cooperative agreement with the National Science Foundation. This project used data obtained with the Dark Energy Camera (DECam), which was constructed by the Dark Energy Survey (DES) collaboration. Funding for the DES Projects has been provided by the U.S. Department of Energy, the U.S. National Science Foundation, the Ministry of Science and Education of Spain, the Science and Technology Facilities Council of the United Kingdom, the Higher Education Funding Council for England, the National Center for Supercomputing Applications at the University of Illinois at Urbana-Champaign, the Kavli Institute of Cosmological Physics at the University of Chicago, Center for Cosmology and Astro-Particle Physics at the Ohio State University, the Mitchell Institute for Fundamental Physics and Astronomy at Texas A\&M University, Financiadora de Estudos e Projetos, Fundacao Carlos Chagas Filho de Amparo, Financiadora de Estudos e Projetos, Fundacao Carlos Chagas Filho de Amparo a Pesquisa do Estado do Rio de Janeiro, Conselho Nacional de Desenvolvimento Cientifico e Tecnologico and the Ministerio da Ciencia, Tecnologia e Inovacao, the Deutsche Forschungsgemeinschaft and the Collaborating Institutions in the Dark Energy Survey. The Collaborating Institutions are Argonne National Laboratory, the University of California at Santa Cruz, the University of Cambridge, Centro de Investigaciones Energeticas, Medioambientales y Tecnologicas-Madrid, the University of Chicago, University College London, the DES-Brazil Consortium, the University of Edinburgh, the Eidgenossische Technische Hochschule (ETH) Zurich, Fermi National Accelerator Laboratory, the University of Illinois at Urbana-Champaign, the Institut de Ciencies de l’Espai (IEEC/CSIC), the Institut de Fisica d’Altes Energies, Lawrence Berkeley National Laboratory, the Ludwig Maximilians Universitat Munchen and the associated Excellence Cluster Universe, the University of Michigan, NSF’s NOIRLab, the University of Nottingham, the Ohio State University, the University of Pennsylvania, the University of Portsmouth, SLAC National Accelerator Laboratory, Stanford University, the University of Sussex, and Texas A\&M University.
The Legacy Surveys imaging of the DESI footprint is supported by the Director, Office of Science, Office of High Energy Physics of the U.S. Department of Energy under Contract No. DE-AC02-05CH1123, by the National Energy Research Scientific Computing Center, a DOE Office of Science User Facility under the same contract; and by the U.S. National Science Foundation, Division of Astronomical Sciences under Contract No. AST-0950945 to NOAO.

Based in part on observations at Cerro Tololo Inter-American Observatory, National Optical Astronomy Observatory, which is operated by the Association of Universities for Research in Astronomy (AURA) under a cooperative agreement with the National Science Foundation.


\section*{Data Availability}

The data underlying this article will be shared on reasonable request to the corresponding author.




\bibliographystyle{mnras}
\bibliography{ref}



\bsp	
\label{lastpage}
\end{document}